\documentclass[12pt, epsfig]{article}                            
\usepackage{epsfig}       
\usepackage{amssymb}    
\usepackage{amsfonts} 
\usepackage[square]{natbib}
\usepackage{graphicx}
\usepackage{dcolumn}
\usepackage{bm}
\usepackage{amsmath}
\usepackage{amsxtra}
\usepackage{amstext}
\usepackage{latexsym}
\usepackage{palatino}
\usepackage[latin1]{inputenc}
\usepackage{epsf}
\usepackage{amsmath,amssymb}
\usepackage{latexsym}
\usepackage{calc}
\usepackage{multicol}
\usepackage{fontenc}
 \newskip\humongous \humongous=0pt plus 1000pt minus 1000pt

\newif\ifdtup

\catcode`\@=11
 
\catcode`\@=11
 
\def\section{\@startsection{section}{1}{\z@}{3.5ex plus 1ex minus
   .2ex}{2.3ex plus .2ex}{\large\bf}}
 
\def\thesection{\arabic{section}}    
\def\thesubsection{\arabic{section}.\arabic{subsection}}

\def\appendix{\setcounter{section}{0}
 \def\thesection{Appendix \Alph{section}}
 \def\thesubsection{\Alph{section}.\arabic{subsection}}
 \def\theequation{\Alph{section}.\arabic{equation}}}

\newcommand{\beq}{\begin{equation}}
\newcommand{\eeq}{\end{equation}}
\newcommand{\bea}{\begin{eqnarray}}
\newcommand{\eea}{\end{eqnarray}}
\newcommand{\beas}{\begin{eqnarray*}}
\newcommand{\eeas}{\end{eqnarray*}}

\newcommand{\bquo}{\begin{quote}}
\newcommand{\enqu}{\end{quote}}

\newcommand{\hsp}{,\hspace{.5cm}}

\begin{document}

\begin{titlepage}
\font\cmss=cmss10 \font\cmsss=cmss10 at 7pt
\leftline{\tt hep-th/0703297}

\vskip -0.5cm
\rightline{\small{\tt ULB-TH/07-11}}

\def\thefootnote{\fnsymbol{footnote}}
\begin{center}
{\Large {\bf Minimum Rate of Dissipation Principle}}\\
{\Large {\bf and Linear Corrections to the Onsager Matrix}}

\vspace{.8cm}

{\large Jarah Evslin and Giorgio Sonnino}

\vspace{.4cm}

{\it 
EURATOM - Belgian State Fusion Association 

Free University of Brussels (U.L.B.), Blvd du Triomphe 

Campus de la Plaine, C.P. 231, Building  NO

Brussels, B-1050, BELGIUM}\\

\vspace{2.1cm}

{\large\bf Abstract}

\end{center}
\vspace{.1cm}

The Minimum Rate of Dissipation Principle (MRDP) affirms that, for time-independent boundary conditions, a thermodynamic system evolves towards a steady-state with the least possible dissipation. In this note, examples of diffusion processes of two solutes in an isothermal system are analyzed in detail. In particular, we consider the relaxation of the system when the metric tensor (i.e. the Onsager matrix) is constant and when the Onsager coefficients weakly depend on the spatial derivatives of the concentrations. We show that, to leading order, during the relaxation towards a steady-state, the system traces out a geodesic in the space of thermodynamic configurations, in accordance with the MRDP.

\vspace{1cm}


\end{titlepage}
\noindent\section{\large\bf Introduction}
\vskip0.5truecm
The {\it Minimum Rate of Dissipation Principle} (MRDP) describes general thermodynamic systems relaxing towards stationary non-equilibrium states (see refs \cite{sonnino1}-\cite{sonnino2}). An interesting application of this principle is to the study of magnetically confined plasmas relaxing  towards a steady state in the highly collisional regime (Pfirsch-Schl{\"{u}}ter regime) and in the low-collisional regimes (the Banana and Plateau regimes). However, before accomplishing this task, we will first test the validity of this principle by analyzing the relaxation processes of several simpler thermodynamic systems. A good experimental test would be to analyze relaxations of solar storms through satellite experiments and check whether the MRDP is really satisfied during the evolution. Theoretical checks can be performed by analyzing the relaxations of thermodynamic systems towards a steady-state through the constitutive evolution equations. This note is devoted to the study of two concrete examples of relaxations. 

Let us now briefly describe the relaxation of a thermodynamic system to a steady-state. In the thermodynamic field theory (TFT) description of a thermodynamic system a homogeneous configuration corresponds to a point x in the thermodynamic space $M$. The corresponding thermodynamic forces, which are completely determined by the configuration x, are assembled into the vector $U$ defined as
\begin{equation}\label{I0} 
\textup{exp}_x(U)=\bar\gamma_{U}(t_f)=0. 
\end{equation} 
Here $\bar\gamma_{U}$ is a shortest path from $x$ to the point $0$, which corresponds to the equilibrium configuration. The exponential map $\textup{exp}_x$ in Eq.~(\ref{I0}) associates to every tangent vector $V$ in $TM_x$ a point $y$ in $M$ which lies a distance $|V|$ along the unique geodesic which is parallel to $V$ at $x$ (see for example ref. \cite{katok}). The system begins at $x$ at time $0$ and reaches the state $y=\gamma(t_f)$ at a finite time $t_f$.

The {\it thermodynamic flows} $J_{\mu}$, which are dual to the thermodynamic forces $U^{\mu}$, and defined to be 
\begin{equation}\label{vt3} 
J_{\mu}=\lambda_{\mu\nu}U^{\nu}=(g_{\mu\nu}+f_{\mu\nu})U^{\nu} 
\end{equation} 
\noindent where $\lambda_{\mu\nu}$ is an asymmetric tensor. $g$, the symmetric part of $\lambda$, is the Riemannian metric on $M$ and the skew-symmetric part is $f_{\mu\nu}$. The {\it entropy production} $\sigma$ at the point $x$ is defined to be 
\begin{equation}\label{vt4} 
\sigma=(g_{\mu\nu}+f_{\mu\nu})U^{\mu}U^{\nu}=g_{\mu\nu}U^{\mu}U^{\nu}\geq 0 
\end{equation} 
\noindent where the inequality corresponds to the second law of thermodynamics. As the Riemannian metric $g_{\mu\nu}$ is positive definite, this inequality is always satisfied. If the point $x$ is not a steady-state then the system will evolve.  We consider the process known as relaxation in which the initial velocity vanishes and the system relaxes to a steady-state $y$. 

The TFT description of relaxation mainly rests upon the following postulate: The manifold (M,g) is a Riemannian manifold with metric $g$.
The Minimum Rate of Dissipation Principle establishes that during the process of relaxation the configuration traces out a geodesic $\gamma$ in the thermodynamic space.  In Refs.~\cite{sonnino1}-\cite{sonnino2} the geodesic property was referred to as the shortest path principle. It allows one to write
\begin{equation}\label{I1}
x=\gamma (t_i),\quad y=\gamma (t_f)=exp_x(V)
\end{equation}
\noindent where $V$ is a tangent vector to $\gamma$ at $x$ whose norm is the geodesic distance from $x$ to $y$.  The system begins at $x$ at time $t_i$ and reaches the steady-state $y$ at the time $t_f$ . The process of relaxation begins as soon as the system is released, that is, immediately after time $t_i$.  In support of this conjecture, we have argued that $V$ automatically satisfies a manifestly covariant form of the Minimum Entropy Production Theorem (valid when the system is near equilibrium), the Glansdorff-Prigogine Universal Criterion of Evolution and the Minimum Rate of Dissipation Principle. 

In general, the macroscopic description of thermodynamic systems gives rise to state variables that depend continuously on the spatial coordinates. The Brussels school results mentioned above allow us to treat spatially extended thermodynamic systems by constructing the following fiber bundle: the ordinary space $R^3$ is combined with the thermodynamic space attached to each point in $R^3$ (this approach can been found in refs~\cite{sonnino3}-\cite{sonnino5}). The ordinary space $R^3$ is the base manifold and the thermodynamic space is the fiber. The G-P theorem, obeyed on each geodesic, takes the integral form
\begin{equation}\label{I2}
\int_{\Omega}J_\mu\frac{dV^\mu}{dt}dv
\end{equation}
 \noindent where $dv$ denotes the volume element and the integration is over the entire space $\Omega$ occupied by the system. 

In this paper the relaxation of a distributed system is illustrated by analyzing two concrete examples. Section~\ref{Onsager} describes the relaxation of a chemical system when the metric is flat (i.e., the Onsager matrix is constant). Then in section~\ref{nonlin} the linear (Onsager) transport coefficients are corrected by terms which linearly depend on the thermodynamic forces (i.e., the Onsager transport coefficients depends on the derivatives of the chemical concentrations). We will see that the Minimum Rate of Dissipation Principle is satisfied in both of these cases.
\vskip 0.5truecm

\noindent\section{\large\bf  Diffusion of Two Solutes in an Isothermal System with Constant Diffusion Coefficients} \label{Onsager}
\vskip 0.5truecm
In this section we analyze in detail a simple example of relaxation: the diffusion of two solutes in an isothermal system when the Onsager matrix is constant. The evolution of the mass fractions $c_i$ is described by the differential equations \cite{onsager}-\cite{onsager2} 
\begin{eqnarray}\label{mdp19}
&&\frac{\partial c_1}{\partial t}=D_{11}\bigtriangleup c_1+D_{12}\bigtriangleup c_2\nonumber\\
&&\frac{\partial c_2}{\partial t}=D_{12}\bigtriangleup c_1+D_{22}\bigtriangleup c_2
\end{eqnarray}
where $\bigtriangleup$ is the Laplacian operator. Eqs~(\ref{mdp19}) are quite general. They describe a direct response of the plasma to the presence of a thermodynamic force. A radial pressure gradient produces a radial particle flux (Fick's law). The corresponding coefficient is the radial electron (ion) radial diffusion coefficient (i.e., {\it the loss of matter}). However, Eqs~(\ref{mdp19}) are also valid for describing general diffusion experiments, which occur {\it e.g.} in isomer mixtures or in isotopic mixtures (with not too different molecular masses) and mass diluted systems ({\it i.e.} $c_1\ll 1$ and $c_2\simeq 1$). 

In this section we will consider the Onsager regime, in which the diffusion coefficients $D_{ij}$ are uniform.  In this case Eqs~(\ref{mdp19}) can be diagonalized and rewritten in terms of the solutes that diffuse independently 
\begin{eqnarray}\label{mdp20}
&&\frac{\partial {\bar c}_1}{\partial t}=D_{1}\bigtriangleup {\bar c}_1\nonumber\\
&&\frac{\partial {\bar c}_2}{\partial t}=D_{2}\bigtriangleup {\bar c}_2
\end{eqnarray}
\noindent where the ${\bar c}_i$ are linear combinations of the mass fractions $c_i$. 

Let us now suppose that our system is infinite and satisfies the initial conditions 
\begin{eqnarray}\label{mdp21}
&&{\bar c}_1({\mathbf r},t=0)=N_1\delta (\mathbf{r})\nonumber\\
&&{\bar c}_2({\mathbf r},t=0)=N_2\delta (\mathbf{r})
\end{eqnarray}
\noindent where the $N_i$ indicate the total number of particles of species $i$. We will now analyze the relaxation of this system and verify the validity of the Minimum Rate of Dissipation Principle. It is easy to check that the solutions of Eq.~(\ref{mdp20}) with initial conditions Eq.~(\ref{mdp21}) can be cast into the form
\begin{eqnarray}\label{mdp22}
&&{\bar c}_1({\mathbf r},t)=\frac{N_1}{(4\pi D_1t)^{3/2}}\exp\Bigl(-\frac{r^2}{4D_1t}\Bigr)\nonumber\\
&&{\bar c}_2({\mathbf r},t)=\frac{N_2}{(4\pi D_2t)^{3/2}}\exp\Bigl(-\frac{r^2}{4D_2t}\Bigr).
\end{eqnarray}
Notice that at each moment $t$ the mass fractions are Gaussian distributions centered at the origin with one standard deviation $\hat\sigma_i$ equal to $\sqrt{2D_it}$.  The origin follows a geodesic in the space of mass fractions; however, we are interested in the space of thermodynamic forces and we will see that the forces always vanish at the origin in this example.

To understand the sense in which the evolution is a geodesic in a space of thermodynamic forces, it will be necessary to follow the relaxation of a point at the time-dependent radius
\beq
r_{\rho}(t)=\rho\sqrt{2Dt}\hsp D\equiv\Bigl(\frac{D_1D_2}{D_2-D_1}\Bigr)
\eeq
which lies a fixed number of standard deviations from the origin.  We will see that the thermodynamic forces at $r_\rho$ follow a geodesic for every fixed value of $\rho$.  One might object that we should instead be following a point at a fixed value of $r$; however, we will see that such points begin already relaxed, then they are excited, and then they relax again.  Thus the shortest path principal, which applies only to the phenomenon of relaxation, does not apply to the value of a field at a fixed value of $r$ in the present example.

The thermodynamic forces are defined to be the gradients of the mass fractions.  In particular their radial components are
\begin{eqnarray}\label{mdp23}
&&U^1(r,t)=\frac{d{\bar c}_1}{d r}=- \frac{r N_1}{16{\pi}^{3/2} (D_1t)^{5/2}}\exp\Bigl(-\frac{r^2}{4D_1t}\Bigr)\nonumber\\
&&U^2(r,t)=\frac{d{\bar c}_2}{d r}=- \frac{r N_2}{16{\pi}^{3/2} (D_2t)^{5/2}}\exp\Bigl(-\frac{r^2}{4D_2t}\Bigr)
\end{eqnarray}
\noindent and so they are related by 
\begin{equation}
U^1(r,t)=\frac{N_1}{N_2}\Bigl(\frac{D_2}{D_1}\Bigr)^{5/2}\exp\Bigl(-\frac{r^2}{4Dt}\Bigr)U^2(r,t). \label{urel}
\end{equation}
\noindent At $r=r_\rho$, Eq.~(\ref{urel}) reduces to
\begin{equation}\label{mdp24}
U^1(r_\rho,t)=\frac{N_1}{N_2}\Bigl(\frac{D_2}{D_1}\Bigr)^{5/2}\exp\Bigl(-\frac{\rho^2}{2}\Bigr)U^2(r,t).
\end{equation}
In particular the ratio of the forces $U^1$ and $U^2$ is independent of $t$ and so, for each value of $\rho$, $U^1(t)$ and $U^2(t)$ lie along a line in the two-dimensional space of forces.  

We are considering the Onsager regime, and so the metric is flat and the line is a geodesic.  Thus at each $r_\rho$ the system traces out a geodesic from a $\rho$-dependent excited state to the stationary state $0$, which in this case is also the equilibrium state. In the Onsager regime the Glansdorff-Prigogine quantity is just the rate at which the forces approaches equilibrium with respect to the affine parameter, which is maximized when the velocity vector points towards equilibrium, as it always does for a straight line ending at equilibrium. Thus the minimum rate of dissipation principle is satisfied, as it must be for any geodesic relaxation according to the general arguments reported in refs (\cite{sonnino1}) and (\cite{sonnino2}).
\vskip 0.5truecm
\noindent\section{\large\bf Diffusion Processes of a Thermodynamic System in which the Metric Tensor is a Small Perturbation of the Onsager Matrix} \label{nonlin}
\vskip 0.5truecm
\noindent We shall now analyze another example: the relaxation of a system when the Onsager coefficients (weakly) depend on the derivatives of the chemical concentrations. We shall show that the minimum rate of dissipation principle is satisfied in this case as well. For simplicity we will consider a system in which the concentrations depend on a single spatial direction, with coordinate $x$.  

We now reconsider the diffusion process examined in the previous section, adding a small perturbation to the metric tensor $g_{\mu\nu}$ with respect to the Onsager tensor $L_{\mu\nu}$
\begin{equation}\label{D1}
g_{\mu\nu}=L_{\mu\nu}+A^\kappa_{\mu\nu}c_\kappa^\prime
\end{equation}
\noindent where $L$ is diagonal, the symbol ${}^\prime$ is the derivative in the $x$ direction and $A^\kappa_{\mu\nu}=A^\kappa_{\nu\mu}$ are considered to be small constant parameters. 

The relaxation processes can be analyzed by solving the following evolution equations for infinite systems 
\begin{eqnarray}\label{D2}
&&\frac{\partial c_1}{\partial t}=-\nabla\cdot{\bf J_1}\qquad\quad {\bf J_1}=g_{11}\nabla c_1+g_{12}\nabla c_2\nonumber\\
&&\frac{\partial c_2}{\partial t}=-\nabla\cdot{\bf J_2}\qquad\quad {\bf J_2}=g_{12}\nabla c_1+g_{22}\nabla c_2.
\end{eqnarray}
\noindent Notice that the structure of these equations is quite general. For example Eq~(\ref{mdp19}) describes diffusion, as well as the thermal processes of Ref.~\cite{degroot}. In this last case, a radial electron (ion) temperature gradient produces a radial electron (ion) heat flux (Fourier's law). The corresponding coefficient is the {\it radial electron (ion) thermal conductivity}. As we are considering uni-dimensional systems, the gradient operator is just $\nabla=(\frac{\partial}{\partial x},0,0)$ and the evolution equations take the form
\begin{eqnarray}\label{D3}
&&{\dot c}_1=D_1c^{''}_1+{\bar A}_1c_1'c_1^{''}+{\bar A}_2c_2'c_2^{''}+{\bar A}_3(c_1'c_2{''}+c_1^{''}c_2')\nonumber\\
&&{\dot c}_2=D_2c^{''}_2+{\bar A}_4c_1'c_1^{''}+{\bar A}_5c_2'c_2^{''}+{\bar A}_6(c_1'c_2{''}+c_1^{''}c_2')
\end{eqnarray}
\noindent where the {\it dot} and {\it prime} indicate the time and the spatial derivatives respectively and $D_k=L_{kk}$ for $k=(1,2)$. Moreover
\begin{equation}\label{D4}
\!{\bar A}_1=2A^1_{11}\ ;\ {\bar A}_2=2A^2_{12}\ ;\ {\bar A}_3=A^2_{11}+A^1_{12}\ ;\ {\bar A}_4=2A^1_{12}\ ;\ {\bar A}_5=2A^1_{22}\ ;\ {\bar A}_6=A^2_{12}+A^1_{22}.
\end{equation}
\noindent At the zeroth order in $A$, we have to solve the equations
\begin{eqnarray}\label{D5}
&&{\dot c}_{10}=D_1c_{10}^{''}\nonumber\\
&&{\dot c}_{20}=D_2c_{20}^{''}
\end{eqnarray}
\noindent with the boundary conditions
\begin{eqnarray}\label{D6}
&&c_{10}({\mathbf r},t=0)=N_1\delta (\mathbf{r})\nonumber\\
&&c_{20}({\mathbf r},t=0)=N_2\delta (\mathbf{r}).
\end{eqnarray}
\noindent We find 
\begin{eqnarray}\label{D7}
&&c_{10}=\frac{N_1}{(4\pi D_1t)^{1/2}}\exp\Bigl(\frac{-x^2}{4D_1t}\Bigr)\nonumber\\
&&c_{20}=\frac{N_2}{(4\pi D_2t)^{1/2}}\exp\Bigl(\frac{-x^2}{4D_2t}\Bigr).
\end{eqnarray}
\noindent As we shall see later on, our calculations are greatly simplified if, instead of coordinates $(x,t)$, we use coordinates $(x,\alpha)$ defined as 
\begin{equation}\label{D8}
x=\alpha\sigma,\quad
t=\frac{\sigma^2}{4D}\quad , \quad D\equiv\frac{D_1D_2}{D_2-D_1}
\end{equation}
\noindent so that $\alpha=x/\sqrt{4Dt}$.  Eqs~(\ref{D7}) take then the form
\begin{eqnarray}\label{D9}
&&c_{10}=\frac{\alpha\sqrt{{\tilde D_1}}N_1}{\sqrt{\pi}x}\exp(-{\tilde D}_1\alpha^2)=\frac{\mu_1}{x}\nonumber\\
&&c_{20}=\frac{\alpha\sqrt{{\tilde D_2}}N_2}{\sqrt{\pi}x}\exp(-{\tilde D}_2\alpha^2)=\frac{\mu_2}{x}\
\end{eqnarray}
\noindent where
\begin{equation}\label{D10}
\mu_k=\alpha {\tilde N}_k\exp (-{\tilde D}_k\alpha^2),\quad {\tilde N}_k=\frac{\sqrt{{\tilde D}_k}N_k}{\sqrt{\pi}},\quad{\tilde D}_k=\frac{D}{D_k},\quad {\rm with}\quad (k=1,2).
\end{equation}

Using
\begin{eqnarray}\label{D11}
&&{\dot\alpha}|_{x \rm\ const}=\frac{-x}{4\sqrt{Dt^3}}=-\frac{2\alpha^3D}{x^2}\nonumber\\
&&\alpha'|_{t \rm\ const}=\frac{1}{\sigma}=\frac{\alpha}{x}
\end{eqnarray}
\noindent and expanding the evolution equations~(\ref{D3}) in $A$ by inserting (\ref{D9}) one finds at leading order
\begin{equation}\label{D12}
{\dot c}_k=D_kc_k^{''}-8A^\lambda_{k\nu}\frac{{\tilde N}_\lambda{\tilde N}_\nu{\tilde D}_\lambda{\tilde D}_\nu\alpha^6}{x^5}\bigl(\alpha^2({\tilde D}_\lambda+{\tilde D}_\nu)-1\bigr)\exp\bigl(-({\tilde D}_\lambda+{\tilde D}_\nu)\alpha^2\bigr)
\end{equation}
\noindent where summation over indexes $\lambda$ and $\nu$ (with $\lambda$ and $\nu$ taking the values $1$ and $2$) is implicitly understood. It is not difficult to check that solution of Eq.~(\ref{D12}) is
\begin{equation}\label{D13}
c_k=c_{0k}\frac{\alpha}{x}\exp(-{\tilde D}_k\alpha^2)+\epsilon_{k\nu}^\lambda{\tilde N}_\lambda{\tilde N}_\nu{\tilde D}_\lambda{\tilde D}_\nu\frac{\alpha^4}{x^3}\exp\bigl(-({\tilde D}_\lambda+{\tilde D}_\nu)\alpha^2\bigr)
\end{equation}
\noindent where
\begin{eqnarray}\label{D14}
&&c_{0k}=\frac{\sqrt{{\tilde D}_k}N_k}{\sqrt{\pi}}\nonumber\\
&&\epsilon^\lambda_{k\nu}=\frac{A^\lambda_{k\nu}}{{\tilde D}_\lambda+{\tilde D}_\nu-{\tilde D}_k}.
\end{eqnarray}

We will now argue that, during the relaxation, the thermodynamic system evolves towards a steady-state along a geodesic. Recall that the thermodynamic forces are defined to be the spatial derivative of the mass fractions. As is easily checked, to first order in $\epsilon$, the derivatives of the affine parameter $\tau$ and of thermodynamic forces, with respect to time $t$, behave as 
\begin{eqnarray}\label{D15}
&&\frac{d\tau}{dt}\sim\frac{1}{\alpha t^2}+O(\epsilon^2)\nonumber\\
&&\frac{dc'}{dt}\sim\frac{1}{\alpha t^2}+\epsilon\frac{1}{t^3}+O(\epsilon^2).
\end{eqnarray}
\noindent On the other hand, to first order in $\epsilon$, we also have 
\begin{eqnarray}\label{D16}
&&
\frac{d^2c'}{d\tau^2}\sim\frac{d^2c'}{dt^2}\Bigl(\frac{dt}{d\tau}\Bigr)^2\sim\Bigl[\frac{d^2}{dt^2}\Bigl(\frac{1}{t^2}\Bigr)\Bigr](\alpha t^2)^2+O(\epsilon^2)\sim O(\alpha^2)+O(\epsilon^2)
\nonumber\\
&&\Bigl(\frac{dc'}{d\tau}\Bigr)^2\sim\Bigl(\frac{d}{d\tau}\frac{1}{t^2}\Bigr)^2+O(\epsilon^2)\sim\Bigl(\frac{1}{t^3}\Bigr)^2(\alpha t^2)^2=O(\alpha^2)+O(\epsilon^2)
\end{eqnarray}
\noindent which is consistent with the geodesic equation, confirming that the evolution of the system traces out a geodesic in the space of thermodynamic configurations. Thus the minimum rate of dissipation principle is satisfied, as it must be for any geodesic relaxation as was demonstrated in refs (\cite{sonnino1}) and (\cite{sonnino2}). 

\noindent\section{\large\bf Conclusions and Perspectives}\label{conclusions}
\vskip0.5truecm
The {\it Minimum Rate of Dissipation Principle} (MRDP) provides predictions for general thermodynamic systems relaxing towards stationary non-equilibrium states (see refs \cite{sonnino1} and \cite{sonnino2}).  In principle, this task can be accomplished only after having set up a formalism where all variables are free and the structure of the fiber bundle is completely determined. In general, the macroscopic description of thermodynamic systems gives rise to state variables that depend continuously on space coordinates. In this case, the thermodynamic forces possess values associated to each spatial point. To treat spatially extended thermodynamic systems, we have to construct a fiber bundle by combining the ordinary space $B=R^3$ with the thermodynamic space F attached to each point in $R^3$. In this and in previous papers, we have treated non-homogenous thermodynamic systems supposing that the fiber is trivial i.e., the fiber bundle is simply the product space $B\times F$. This is, however, a strong assumption, in fact there are a wide variety of examples where the spaces cannot be described as trivial fiber bundles. 

In this paper we have checked the validity of this principle by analyzing the relaxation processes of several thermodynamic systems, assuming that the fiber is trivial. In particular we analyzed the relaxation of two concrete examples:  systems where the metric tensor (the Onsager matrix) is independent of the thermodynamic forces and the case where the Onsager matrix weakly depends on the gradients of the concentrations. The structure of these equations is quite general: they can be used for describing diffusion, as well as thermal processes. A typical example is a radial electron (ion) temperature gradient, which produces a radial electron (ion) heat flux (Fourier's law). The corresponding coefficient is the {\it radial electron (ion) thermal conductivity}.  We proved that to leading order the evolution of the system traces out a geodesic in the space of thermodynamic configurations. From ref.~\cite{sonnino1}, we then have also shown that in these cases the systems evolve towards a steady-state with the least possible dissipation (the {\it Minimum Rate of Dissipation Principle}). 

\noindent 
\vskip 0.5truecm
\noindent\section{\large\bf Acknowledgments}\label{acknowledgments}
\noindent One of us (GS) would like to express his sincere gratitude to his hierarchy at the European Commission and to the members of the EURATOM Belgian State Fusion Association. The work of JE is partially supported by IISN - Belgium (convention 4.4505.86), by the ``Interuniversity Attraction Poles Programme -- Belgian Science Policy'' and by the European Commission RTN program HPRN-CT-00131, in which he is associated to K. U. Leuven. 
\vskip0.5truecm


\end{document}